\documentclass[prl,aps,footinbib,twocolumn,showpacs]{revtex4}
\usepackage{amssymb}
\usepackage{amssymb}
\usepackage{graphicx}
\usepackage{amsmath}
\usepackage{times}
\usepackage{color}
\usepackage{subfigure}
\usepackage{setspace}
\usepackage{bm}

\begin{document}

\title{Many-body treatment of  the collisional frequency shift in fermionic atoms }

\author{ A.M.~Rey$^1$, A.V.~Gorshkov$^2$, C.~Rubbo$^1$ }
\affiliation{$^1$JILA, NIST and  Department of Physics, University of Colorado
Boulder, CO 80309,}
\affiliation{$^{2}$Physics Department, Harvard University, Cambridge, MA, 02138}


\begin{abstract}
Recent clock experiments have measured density-dependent  frequency shifts in polarized fermionic alkaline-earth atoms using ${}^1S_0$-${}^3P_0$ Rabi spectroscopy. Here we provide a first-principles non-equilibrium  theoretical description of the interaction frequency shifts starting  from the microscopic many-body Hamiltonian. Our formalism describes the dependence of the frequency shift on  excitation inhomogeneity, interactions, and many-body dynamics, provides a fundamental  understanding of  the effects of the  measurement process, and explains the observed density shift  data. We also propose a method
to measure the second of the two 
${}^1S_0$-${}^3P_0$ scattering lengths, whose knowledge is essential for 
quantum information processing and quantum simulation applications.
\end{abstract}

\date{\today }
\maketitle

Experimental efforts in cooling, trapping, and manipulating alkaline-earth-like atoms such as Sr and Yb have led to unprecedented developments in optical clocks   based on the ${}^1S_0$-${}^3P_0$ transition \cite{Ye2008,Ludlow2008,Campbell2009,Akatsuka2008,Takamoto2009,Fukuhara2007b1,Fukuhara2007b,Lisdat2009, Barber2008,Lemke2009}. 
While the interrogation of a large number of atoms enhances the sensitivity of these clocks, the accompanying interatomic interactions degrade clock precision. Thus the understanding of these interactions is crucial for precision spectroscopy. Fermionic alkaline-earth atoms
 have also started to attract considerable theoretical attention in the context of quantum information processing \cite{Reichenbach2007, Hayes2007,
Daley2008, Gorshkov2009} and quantum simulation \cite{Gorshkov20092}, 
applications requiring the knowledge of both ${}^1S_0$-${}^3P_0$ scattering lengths. 
Here we present  a many-body formulation, which, on the one hand, allows us to understand the nature of the collisional frequency shift (CFS) measured in  a recent experiment \cite{Campbell2009}, and on the other, allows us to propose a way 
to measure the remaining 
${}^1S_0$-${}^3P_0$ scattering length, which was not probed in Ref.~\cite{Campbell2009}.
Our model helps to clarify  the role of excitation inhomogeneities, dynamics, and interactions  in fermionic clock experiments \cite{Gupta2003,Zwierlein2003,Campbell2009,Blatt2009}.

Clock experiments based on  Rabi interrogation start with a  nuclear-spin-polarized sample of atoms prepared (for consistency with Ref.~\cite{Campbell2009}, which we aim to model) in an excited state $e$, which is then  transferred     to the ground state $g$ by illuminating the atoms during a time $t_f$ with a probe beam  detuned from the atomic resonance. The CFS $\delta \omega_{eg}$ is inferred by  recording the  final population in $g$ as a function of the detuning and looking for  changes in the corresponding lineshape due to interactions.
So far most  treatments of CFSs   in dilute polarized fermionic gases  away from the unitarity limit were based on a static mean field analysis \cite{Zwierlein2003,Gupta2003,Oktel1999, Oktel2002}. The latter predicts a frequency shift $\delta \omega_{eg}  =\frac{4 \pi \hbar a_{eg}^-}{M} (\rho_g-\rho_e) G_{ge}^{(2)}$, with $G_{ge}^{(2)}$ the two atom correlation function at zero distance, which measures the probability  that two particles are simultaneously detected, $a_{eg}^-$ the s-wave scattering length between the $g$ and $e$ atoms with mass $M$, and  $\rho_{g,e}$ the corresponding atom densities. Here we extend this formulation beyond mean-field and fully account for the many-body dynamics during Rabi interrogation. Our key statements are as follows. 
(i) Motion-induced excitation inhomogeneity can lead to  s-wave CFS even in an initially  polarized ensemble of fermions.  (ii) CFS is sensitive to the time-averaged population difference  between $g$ and $e$ atoms, consistently with the mean field approximation,  and in particular  vanishes when this difference goes to zero ($\pi$ pulse). (iii) For a fixed pulse area,  the CFS approaches zero as $t_f \rightarrow 0$, meaning that, in order to experience  interactions, atoms should have enough time to feel the 
excitation inhomogeneity. (iv)
Measurements of $\delta \omega_{eg} $ done  by locking the interrogation laser at fixed final ground state fraction are very sensitive to the pulse area and strength of interactions. Depending on these parameters, even the  sign of $\delta \omega_{eg} $ can be reversed.

We begin our analysis with the Hamiltonian $\hat H$ describing cold fermionic alkaline-earth atoms illuminated by a linearly polarized laser beam with bare Rabi frequency $\Omega_0$ and  trapped in an external potential $V(\mathbf{r})$ that is the same for $g$ and $e$ (i.e.~at the ``magic wavelength" \cite{Ye2008}). 
Assuming that the atoms are polarized 
in a state with nuclear spin projection 
$m_0$, 
we omit the nuclear spin label and, setting $\hbar = 1$, obtain  \cite{Gorshkov2009,Gorshkov20092}
 \begin{eqnarray}
\hat H = \sum_{\alpha }  \! \int  \!\! d^3  \mathbf{r} \hat \Psi^\dagger_{\alpha }   \left(- \frac{1}{2 M} \nabla^2 + V(\mathbf{r})\right) \hat \Psi_{\alpha }   +u_{eg}^-  \! \int  \!\! d^3 \mathbf{r} \hat \rho_e \hat \rho_g \nonumber  \\
+\omega_0  \! \int  \!\! d^3 \mathbf{r} (\hat \rho_e - \hat \rho_g)
- \frac{ \Omega_0}{2}  \! \int  \!\!  d^3 \mathbf{r} (\hat \Psi^\dagger_{e } e^{-i (\omega_L t- \bm{k} \cdot \bf{ r})} \hat \Psi_{g } + {\rm h.c.}). \label{ham0}
\end{eqnarray}
Here $\hat \Psi_{\alpha }(\mathbf{r})$ is a fermionic field operator at position $\mathbf{r}$ for atoms in electronic  state  $\alpha = g$ ($^1S_0$) or $e$ ($^3P_0$), while $\hat \rho_{\alpha }(\mathbf{r}) =
\hat \Psi^\dagger_{\alpha }(\mathbf{r}) \hat \Psi_{\alpha }(\mathbf{r})$ is the corresponding density operator.   
 Since polarized fermions are in a symmetric nuclear state, their  $s$-wave interactions are characterized by only one   scattering
length $a_{eg}^-$, with the corresponding interaction parameter $u_{eg}^-= 4 \pi \hbar^2 a_{eg}^-/M$, describing collisions between two atoms in the antisymmetric electronic state $|-\rangle =(|ge\rangle
-|eg\rangle)/\sqrt{2}$. 
The laser with frequency $\omega_L$ and wavevector $\bm{k}$ is detuned from the atom transition frequency $\omega_0$ by $\delta=\omega_L-\omega_0$.

As in the experiment of Ref.~\cite{Campbell2009}, 
we assume that most atoms are frozen along one (longitudinal) $z$-direction, leaving in the remaining transverse $x$-$y$ plane an isotropic 2D harmonic oscillator with frequency $\omega_x = \omega_y$. 
We can then write 
$\hat{\Psi}_{\alpha}(\mathbf{r})=\phi^z_0(z)  \sum_{ \bm \nu} \hat{c}_{\alpha \bm \nu} \phi_{\nu_x} (x) \phi_{\nu_y}(y)$, where 
$\phi_\nu^z$ and $\phi_\nu$
are, respectively, the longitudinal and the transverse harmonic oscillator eigenmodes and $\hat{c}^\dagger_{\alpha \bm \nu}$ 
creates a fermion in mode $\bm \nu = (\nu_x, \nu_y)$ and  electronic level $\alpha$. Following Refs.~\cite{Campbell2009, Blatt2009}, we assume that the probe is slightly misaligned from the $z$-direction: $\bm{k} = k_z \hat z + k_x \hat x$ with $|k_x/k_z| \ll 1$. Defining then $\Omega_{\nu,\nu'}= \Omega_0 e^{-\eta_z^2/2} L_0(\eta_z^2) \langle \phi_{\nu}(x)|e^{ i k_x x}|\phi_{\nu'}(x)\rangle$, where $\eta_i=k_i\sqrt{\frac{\hbar}{2 m\omega_i}}\ll 1$ are the  Lamb-Dicke parameters and $L_n$ are Laguerre polynomials \cite{Wineland1979}, laser induced  sideband  transitions can be neglected if
 $ \Omega_{\nu,\nu' \neq \nu}\ll \omega_x$.  In this regime, $\hat H$  can be rewritten in the rotating frame as 
\begin{eqnarray}
&&\hat H=- \delta \sum_{\bm \nu} \hat{n}_{e \bm \nu} +\sum_{\bm \nu,\alpha} E_{\bm \nu} \hat{n}_{\alpha \bm \nu}-\sum_{\bm \nu}{\frac{\Omega_{\nu_x}}{2}(\hat{c}^\dag_{g \bm \nu}\hat{c}_{e \bm \nu}+ \rm{h.c})} \nonumber\\&& +u^-_{eg}\sum_{\bm \nu_1 \bm \nu_2 \bm \nu_3 \bm \nu_4}{A_{\bm \nu_1\bm \nu_2 \bm \nu_3 \bm \nu_4}\hat{c}^\dag_{e \bm \nu_1}\hat{c}_{e \bm \nu_2}\hat{c}^\dag_{g \bm \nu_3}\hat{c}_{g \bm \nu_4}},
 \label{many0}
\end{eqnarray} where $A_{\bm \nu_1\bm \nu_2 \bm \nu_3 \bm \nu_4}=\int (\phi^z_0)^4 dz\int \prod_j \phi_{\nu_{j x}} dx \int \prod_j \phi_{\nu_{j y}} dy$, $\hat{n}_{\alpha \bm \nu} =\hat{c}^\dagger_{\alpha \bm \nu} \hat{c}_{\alpha \bm \nu}$,
  $\Omega_{\nu_x} = \Omega_0 L_{\nu_x}(\eta_x^2) L_0(\eta_z^2) e^{-(\eta_x^2 + \eta_z^2)/2}$,  and $E_{\bm \nu}$ are single-particle energies.

Many interaction terms in Eq.~(\ref{many0}) can be ignored provided that the interaction is weaker than $\omega_x$ and that the 2D oscillator is slightly anharmonic as in the experiment of Ref.~\cite{Campbell2009}. Furthermore, we note that unless $\nu_1 = \nu_2$ and $\nu_3 = \nu_4$, $|\int \phi_{\nu_1}^2 \phi_{\nu_3}^2| \gg |\int \prod_j \phi_{\nu_{j}}|$. As a result, the interaction is dominated by the terms where $(\bm \nu_{1},\bm \nu_{3})$ is equal to $(\bm \nu_{2}, \bm \nu_{4})$, $(\nu_{4x},\nu_{2y},\nu_{2x},\nu_{4y})$, $(\nu_{2x},\nu_{4y},\nu_{4x},\nu_{2y})$, or $(\bm \nu_{4},\bm \nu_{2})$, which describe the exchange of modes along neither direction, along $x$, along $y$, and along both directions, respectively. Postponing the study of the terms exchanging modes along one direction only, the remaining terms conserve the number of particles per mode $\bm \nu$.
 Assuming there is only one atom in each of $N$ non-empty modes $\vec{ \bm \nu}= \{\bm \nu_1,\dots, \bm \nu_N\}$, as is the case if all atoms are initially in the same internal state, 
 $\hat H$ can then be reduced to a  spin-$1/2$ model describing these modes:
\begin{eqnarray}
\hat H_{S} = -  \delta \hat{S}_z - \sum_{\bm \nu}{\Omega_{\nu_x} \hat{S}_{x}^{\bm \nu}} -   \sum_{\bm \nu \neq \bm \nu' }{ U_{\bm \nu \bm \nu'} (\vec{\hat S}^{\bm \nu}\cdot \vec{\hat S}^{\bm \nu'} - 1/4)}. \label{many}
\end{eqnarray}
Here $U_{\bm \nu \bm \nu'}=u_{eg}^-A_{\bm \nu \bm \nu \bm \nu' \bm \nu'}$, $\vec{\hat{S}}^{\bm \nu} = \frac{1}{2}\sum_{\alpha,\alpha'}\hat{c}^\dag_{\alpha \bm \nu}\vec{\sigma}_{\alpha \alpha'}\hat{c}_{\alpha' \bm \nu}$, where $\vec \sigma$ are Pauli matrices in the $\left\{e,g\right\}$ basis, $\hat S_{i=x,y,z} = \sum_{\bm \nu} \hat{S}_{i}^{\bm \nu}$, and constant terms were dropped.
  $\hat H_S$ is reminiscent of solid-state spin Hamiltonians, which can also feature long-range interactions and rich nonequilibrium dynamics \cite{Dutt2007}.

The rotational invariance of the interaction term  in $\hat H_S$ ($\propto U_{\bm \nu \bm \nu'}$) is  key to understanding some of the basic features of the model.
The interaction term is  diagonal in the collective angular momentum  basis $|S,M,q\rangle$, satisfying  $\hat{S}^2|S,M,q\rangle=S(S+1)|S,M,q\rangle$ and $\hat{S}_z|S,M,q\rangle=M|S,M,q\rangle$, with $S=0,\dots N/2$ and $-S\leq M\leq S$.
Here the extra label $q$ is required to uniquely specify each state. 
 The fully symmetric (Dicke)
$S=N/2$ states 
do not interact. They are unique and the label $q$ can be omitted for them.

  For a homogeneous excitation, 
  $\Omega_{\nu}=\bar{\Omega}$, 
  the term $\sum_{\bm \nu}{\Omega_{\nu_x} \hat{S}_{x}^{\bm \nu}}$ 
  commutes with $\hat{S}^2$,   the interaction energy is 
  conserved,
  and  no CFS  will be observed provided the initial state is an eigenstate of the interaction or a classical mixture of them, consistent with Ref.~\cite{Zwierlein2003}.
 If the system is prepared  in the Dicke manifold, states with  $S<N/2$ are never populated and
the ground state population evolves collectively as ${N}_g^{(0)}(t,\delta)= N\frac{\bar{\Omega}^2}{\bar{\Omega}^2+\delta^2} \sin^2\left (\frac{t \sqrt{\bar{\Omega}^2+\delta^2}}{2}\right),
$ where the superscript ${}^{(0)}$ indicates a homogeneous excitation. 

In the presence of excitation inhomogeneity, there are two simple limiting cases where  CFSs are absent:
the non-interacting regime and  the strongly interacting regime 
where interactions dominate over Rabi frequency inhomogeneity. The suppression of CFSs  in the latter case is a consequence of the large  energy  gap between states with different $S$, which  brings $S$-changing transitions out of resonance \cite{Rey2008a}. 
In the intermediate interaction  regime, on the contrary, the excitation 
inhomogeneity cannot be ignored. It will transfer atoms  between states with different $S$  generating a net CFS even at zero temperature.

\begin{figure}
\centering
{\includegraphics[width=2.5in]{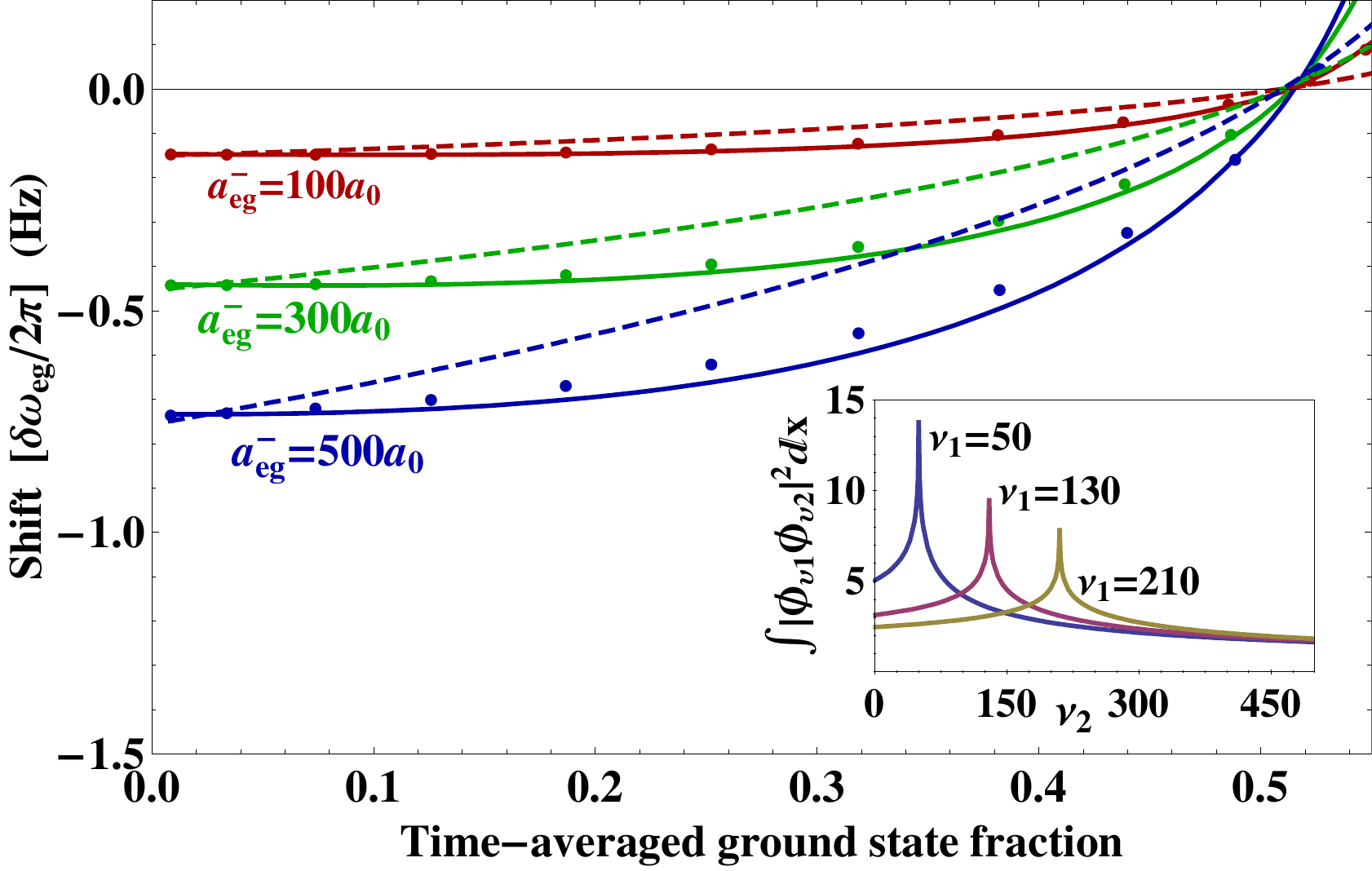}}
\caption{(color online) CFS, $\delta\omega_{eg}/(2\pi)$, at resonant transfer, for  different $a_{eg}^-$ (in Bohr radii, $a_0$). The time-averaged ground state fraction was varied by changing  $t_f$. The solid lines were calculated by thermally averaging $N_g$ computed from Eq.~(\ref{many}), and the dots by a single realization of $\vec{\bm \nu}$ randomly chosen out of those satisfying $\Delta \Omega(\vec{\bm \nu}) =\langle \Delta \Omega\rangle_T$, $\bar{\Omega}(\vec{\bm \nu}) = \langle \bar{\Omega}\rangle_T$, and $\bar{U}(\vec{\bm \nu}) = \langle U\rangle_T$. The dashed lines show Eq.~(\ref{shift}) evaluated at thermally averaged parameters. Here  $N=7$, $T=1\mu$K, $\langle \Delta \Omega \rangle_T/\langle \bar{\Omega}\rangle_T =0.2$ and  $\bar{\Omega}=6\pi/(80\textrm{ ms})$. The inset  shows $\int \phi_{\nu_1}^2(x) \phi_{\nu_2}^2(x) dx$ in  arbitrary units. \label{ResonantFig}}
\end{figure}

In most experiments, e.g.~Ref.~\cite{Campbell2009},  $\delta \omega_{eg}$ is measured  by first locking  the spectroscopy laser  at  two  points, $\delta_{1,2}$, of equal height
in the transition lineshape (equal final ground state fraction under the initial condition of all atoms in state $e$)  and then determining  the change in the mean frequency as the interaction parameters  or density  are varied,
$\delta \omega_{eg} =(\delta_1+\delta_2)/2$. Note that for a homogeneous excitation 
${N}^{(0)}_g(t,\delta) = {N}^{(0)}_g(t,-\delta)$ and therefore  $\delta \omega_{eg} =0$.
Defining $\bar{\Omega}(\vec{\bm \nu})$ to be the mean Rabi frequency over modes $\vec{ \bm \nu}$ and treating $\sum_{\bm \nu}(\Omega_{\nu_{x}}-\bar{\Omega}(\vec{\bm \nu})) \hat S_{x}^{\bm \nu}$ as a perturbation, we write  ${N}_g(t_f,\delta)$ as ${N}_g^{(0)}(t_f,\delta)+ N_g^{(2)}(t_f, \delta)$ (the first order term vanishes), 
Taylor expand it around $\pm \delta_1^{(0)}$,  and 
obtain 
\begin{eqnarray}
\delta \omega_{eg}  \approx \frac{N_g^{(2)}(t_f,-\delta_1^{(0)})-N_g^{(2)}(t_f,\delta_1^{(0)})}{2 \frac{\partial N_g^{(0)}(t_f,\delta)}{\partial\delta}|_{\delta_1^{(0)}}}. \label{geshi}
\end{eqnarray}

To proceed further, we note that $U_{\bm \nu \bm \nu'}$ 
is a slowly varying 
function of $|\nu_{i}-\nu'_{i}|$ ($i = x, y$) [see Fig.~\ref{ResonantFig} (inset)], except within a 
narrow range near $\nu_{i} = \nu'_{i}$. 
Provided $k_B T \gtrsim  N \omega_z$ (which was satisfied in Ref.~\cite{Campbell2009}), the occupied modes $\vec{\bm \nu}$ are sufficiently sparse for the behavior of $U_{\bm \nu_i \bm \nu_j}$ to be dominated by its slowly varying part. 
Therefore, we can approximate $U_{\bm \nu_i \bm \nu_j} \to \bar{U}(\vec{\bm \nu})$. 
Under this approximation, 
the states with $S=N/2-1$, so called spin-wave states, are separated in  energy from the Dicke states by  $\bar{U}(\vec{\bm \nu})N$ and are the only states exited to first order by the vector perturbation operator $\sum_{\bm \nu}(\Omega_{\nu_{x}}-\bar{\Omega}(\vec{\bm \nu})) \hat S_{x}^{\bm \nu}$. 
This allows us to obtain an analytic expression for $\delta \omega_{eg}$ that depends on $\Omega_{\nu_{i x}}$ only through $\bar{\Omega}(\vec{\bm \nu})$ and $\Delta \Omega(\vec{\bm \nu})$,  the root-mean-square Rabi frequency.
This expression is particularly simple  and illuminating when evaluated at resonant population transfer ($\delta_1^{(0)}\to 0$):
\begin{equation}
\small{\delta \omega_{eg}\left|_{{}_{\delta_1^{(0)}\to 0}}\right.=-  \frac{ \Delta\Omega^2 N \bar{U} }{  \bar{\Omega}^2} \frac{  \sin(\mathcal{A}) }{\mathcal{A}} \frac{  \sin[(N\bar{U})  t_f]}{ (N \bar{U})  t_f} f(\bar{\Omega},t_f, N \bar{U})}.\label{shift}
\end{equation}
Here $ \mathcal{A}(\bm \vec{\nu})= \bar{\Omega }(\bm \vec{\nu}) t_f$  is the pulse area and
$f(\bar{\Omega},t_f, N \bar{U})=\frac {1 - \bar{\Omega}/(N \bar{U})  \tan( t_f (N \bar{U})/2)\cot( \mathcal{A}/2)}{ ( 1- (N\bar{U})^2/\bar{\Omega}^2)[4  \sin^2(\mathcal{A}/2)/\mathcal{A}^2 - \sin(\mathcal{A})/\mathcal{A} ]}$. The dependence of  $\bar{U},\bar{\Omega }$, and $\Delta\Omega$ on $\vec{\bm \nu}$ 
is implied.   We now make 
a few  important remarks:
(a) In the limit $t_f \to 0$ ($N_e\to 1$), Eq.~(\ref {shift}) reproduces  the mean-field expression $\delta \omega_{eg} \to - \frac{(\Delta\Omega)^2 N \bar{U}}{\bar{\Omega}^2 } \propto - N \bar{U}
G_{ge}^{(2)}[t_f\to 0,\delta\to 0]$ \cite{Campbell2009, Blatt2009}.
(b)  $\delta \omega_{eg}$  depends on the time-averaged population difference, $ \langle N_g-N_e\rangle_{t_f}=\frac{1}{t_f}\int_{0}^{t_f} (N_g(\tau)-N_e(\tau)) d\tau |_{\delta=0, \Delta \Omega=0} =-\frac{  N \sin(\mathcal{A})}{\mathcal{A}} $ and exactly vanishes at $\mathcal{A}=\pi$
 when $\langle N_g\rangle_{t_f}=\langle N_e\rangle_{t_f}$.
(c) For a fixed pulse area $\mathcal{A}$, as $t_f \rightarrow 0$, 
the frequency shift vanishes as  $\delta \omega_{eg}\to  -(\Delta\Omega t_f)^2 N \bar{U}   \cot[ \mathcal{A} /2]/(2  \mathcal{A})$, implying that in order to experience a CFS atoms  need 
time to feel the 
excitation inhomogeneity.
(d)  At finite times $t_f N\bar{U}\gtrsim 1$ and  $N\bar{U} \gg \bar{\Omega} $,   $\delta \omega_{eg}\propto \frac{ \sin[t_f (N\bar{U}) ]}{(N\bar{U})^2}$ reproducing  the expected  suppression  in the strongly interacting  limit.

So far we have assumed a fixed set of populated modes, $\vec{\bm \nu}$. At finite temperature, expectation values need to be calculated by  averaging  over all possible  combinations of modes $\{ \vec{\bm \nu}\}$
weighted  according to their  Boltzmann factor. However, since  the quantities  $\Delta \Omega(\vec{\bm\nu})$, $\bar{\Omega}(\vec{\bm\nu})$, and $\bar{U}(\vec{\bm\nu})$ are  sharply peaked around their thermal averages, to a good approximation, $\langle N_g\rangle_T$ and thus $\langle \delta \omega_{eg}\rangle_T$ can be calculated by replacing $\Delta \Omega(\vec{\bm\nu}) \to \langle \Delta \Omega\rangle_T$, $\bar{\Omega}(\vec{\bm\nu}) \to \langle \bar{\Omega}\rangle_T$, and $\bar{U}(\vec{\bm \nu}) \to \langle\bar{U}\rangle_T$. Here $\langle \mathcal {O}\rangle_T =\frac{\sum_{\vec{\bm\nu}}\mathcal {O}(\vec{\bm\nu})   e^{-E(\vec{\bm\nu})/(k_B T)} }{\sum_{\vec{\bm\nu}} e^{-E(\vec{\bm\nu})/(k_B T)}}$. The validity of this approximation is demonstrated in Fig.~\ref{ResonantFig}, which also shows that Eq.~(\ref{shift}) is in fair agreement with Eq.~(\ref{many}). 

\begin{figure}
\centering
\includegraphics[width=2.5in]{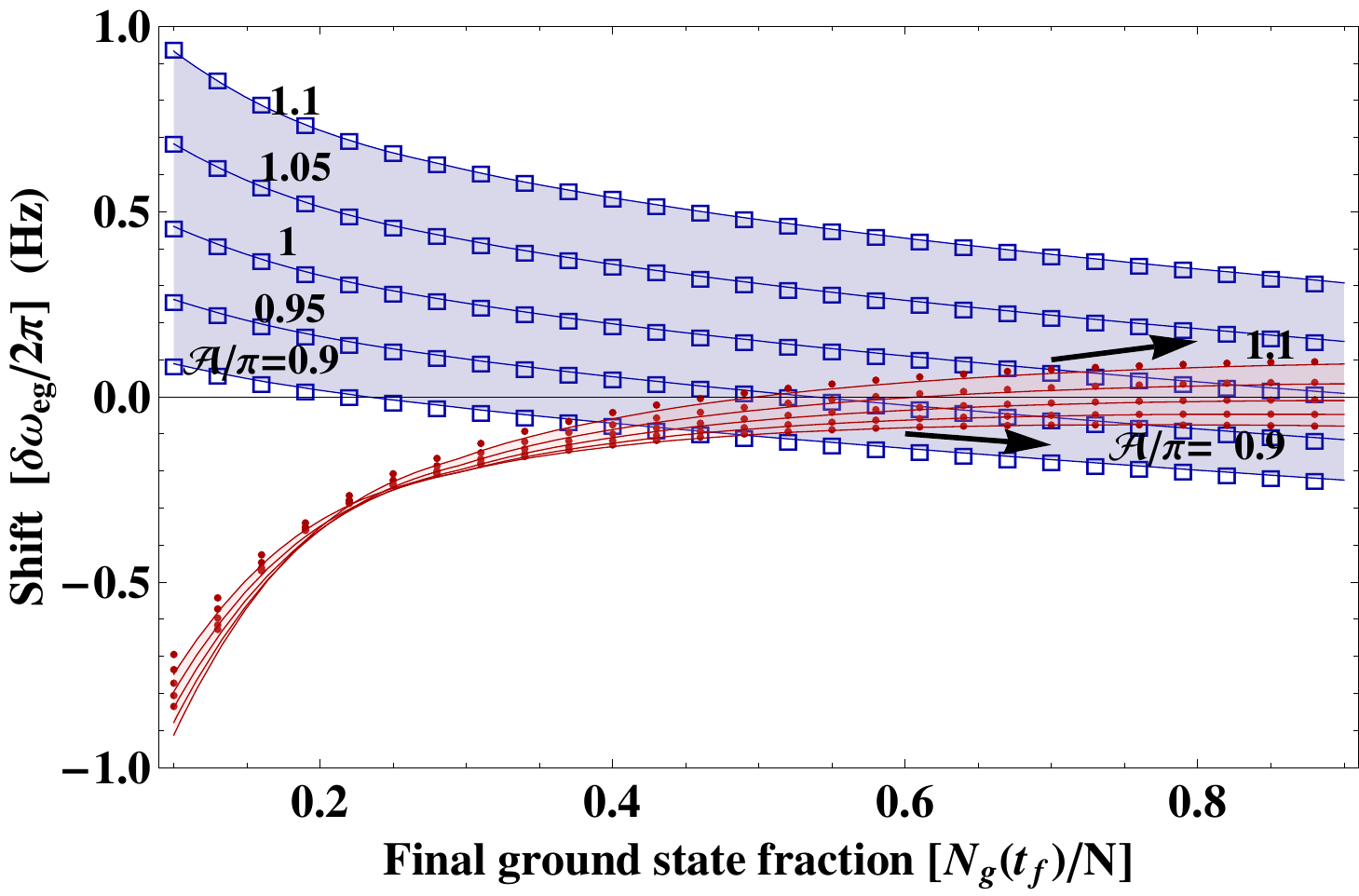}
\caption{(color online) CFS   for different pulse areas, $\mathcal{A}$.
The final ground state  fraction, $N_g(t_f)/N$, was varied by changing the detuning. Here four  spatial modes  at the corners of a square in the $\nu_x$-$\nu_y$ plane,  $ {\bm \nu}_1=(30, 43)$, $ {\bm \nu}_2=(101, 71)$, $ {\bm\nu}_3=(30, 71)$, and ${\bm \nu}_4=(101,43)$, were assumed with $N=2$ atoms occupying ${\bm \nu}_1$ and ${\bm \nu}_2$ at $t=0$. The solid  blue (red) lines, for $\bar {U} N/\bar {\Omega}=1$ $(3.25)$ and $t_f=\pi/\Omega_0= 7$ms, were obtained
using Eq.~(\ref{many0}),  which allows to populate $\bm \nu_3$ and $\bm \nu_4$.
The empty squares (dots) were obtained using  $\hat H_S$ with an effective $\Omega_0^{eff} = \Omega_0/2$ and  $t_f^{eff}=\pi/\Omega_0^{eff}$. \label{OffResonantFig}}
\end{figure}

Experimentally it is hard to measure CFSs close to resonant population transfer due to the small signal-to-noise ratio, and instead the probe  laser is generally locked at a finite detuning \cite{Campbell2009}. Away from  $\delta_1^{(0)}=0$, we have to consider the more general expression given by Eq.~(\ref{geshi}), and an intuitive interpretation is not straightforward.  The interaction induced asymmetry in the lineshape
not only can change the sign of the frequency shift, 
 but, in general, 
makes CFS 
a very sensitive  function of $\langle \mathcal{A} \rangle_T$, $\delta_1^{(0)}$, and  $\langle N\bar{U} \rangle_T$. In particular, the solid lines in Fig.~\ref{OffResonantFig} show the CFS as a function of the ground state fraction at $t_f$, computed using Eq.~(\ref{many0}) for two atoms and four spatial modes.  
For weak repulsive  interactions $0<\langle N \bar{U} \rangle_T\lesssim  \langle \bar{\Omega} \rangle_T$ (blue lines) and  pulse areas greater than $\pi$, $\delta \omega_{eg}>0$ for any detuning and approaches zero only as $\delta_1^{(0)}\to 0$ and $A\to \pi$ (consistently with remark b above). For  $ \mathcal{A}<\pi$, $\delta \omega_{eg}$ 
changes sign as a function of $N_g(t_f)$, 
and the zero crossing point moves  towards 
smaller $N_g(t_f)$
with decreasing pulse area. For stronger interactions 
$\langle N \bar{U} \rangle_T >  \langle \bar{\Omega} \rangle_T>0$ (red lines), while at large $N_g(t_f)$ the sign of $\delta \omega_{eg}$ also depends on whether $\langle \mathcal{A} \rangle_T$ is larger or smaller than $\pi$, at small $N_g(t_f)$  the magnitude of the CFS   becomes less sensitive to pulse area variations and recovers the expected negative sign  for repulsive interactions (since at $t=0$ all atoms are in $e$). All these conclusions are 
 consistent with the  measurements reported  in Ref.~\cite{Campbell2009}
 since  $\sim\!\!10 \%$  variations in $\mathcal{A}$ 
 over the course of a day  could not be excluded.

We now discuss the effect of the terms whose omission in the derivation of Eq.~(\ref{many}) was not justified: the terms exchanging modes along one direction only. As shown in Fig.~\ref{OffResonantFig} for the case of two atoms occupying four modes,  the spin model $\hat H_S$ (dots) reproduces the result of Eq.~(\ref{many0}) (lines) very well provided that we reduce $\Omega_0$ in $\hat H_S$ by a factor of 2: $\Omega_0^{eff} = \Omega_0/2$. This approximate result can be derived analytically by noting that the $\bm \nu_1$-$\bm \nu_2$ singlet ($|-\rangle$) from the spin model is replaced in Eq.~(\ref{many0}) by the symmetric linear combination of the $\bm \nu_1$-$\bm \nu_2$ and the $\bm \nu_3$-$\bm \nu_4$ singlets. 
We have checked that for $N = 3$ and $4$ with $N^2$ modes, $\hat H_S$ with $\Omega_0^{eff} = \alpha \Omega_0$ ($1/2 < \alpha < 1$) also reproduces well the results of Eq.~(\ref{many0}).

\begin{figure}
\centering
{\includegraphics[width=2.8in]{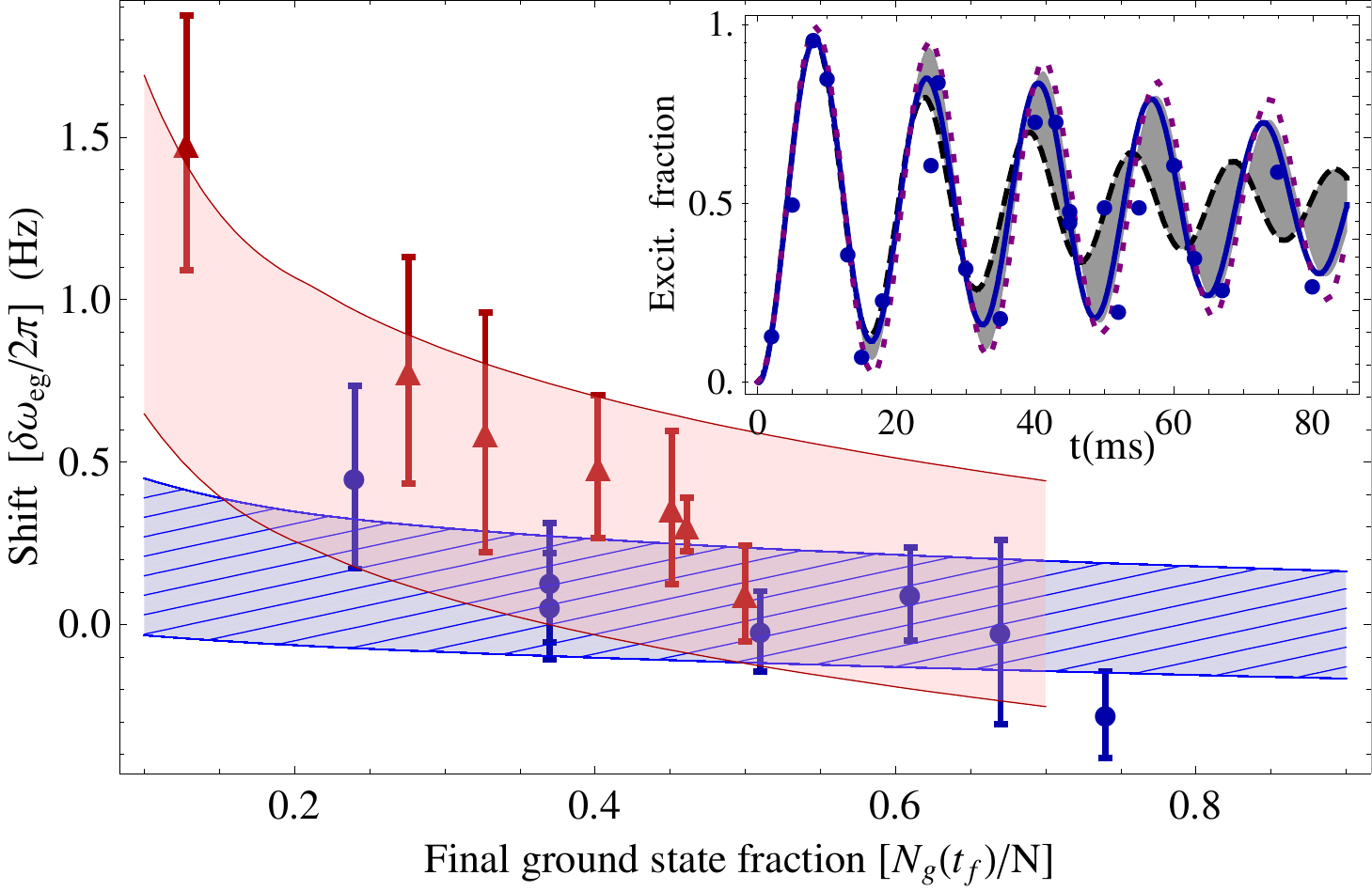}}
\caption{(color online) CFS predicted by our spin model
for   $T=1(3) \mu$K and $\langle \Delta \Omega \rangle_T/\langle \bar {\Omega} \rangle_T=0.15$ ($0.42$), determined from the Rabi flopping curves [see inset]. The blue (pink) shaded 
area shows the uncertainty in  CFS for $T=1(3) \mu$K, assuming a variation in pulse area of $\langle \mathcal A \rangle_T/\pi=1\pm 0.1$.
We used  $\Omega_0^{eff}=4 \pi/ 80  {\rm ms}$, $t_f=\pi/\Omega_0^{eff}$,  $N=15$ [i.e. $\rho\sim 10^{11} {\rm {cm}}^{-3}$], and $ a_{eg}^-=200 a_0$. The circles (triangles)  show the  $T=1(3) \mu$K  experimental  data of Ref.~\cite{Campbell2009} at   $t_f=80$ ms and  $\rho\sim 10^{11} {\rm {cm}}^{-3}$. Inset: The shaded 
area shows Rabi flopping curves calculated at $\langle \Delta \Omega \rangle_T/\langle \bar {\Omega} \rangle_T=0.15$ for   $a_{eg}^-=0-600 a_0$. The dashed black and solid blue  lines correspond to two specific scattering lengths, $a_{eg}^-=0$ and $200 a_0$, respectively. The dotted purple line  is for  $\langle \Delta \Omega \rangle_T/\langle \bar {\Omega} \rangle_T=0.05$ and  $a_{eg}^-=0$, while the blue circles are experimental 
data points from Ref.~\cite{Campbell2009}.\label{ExperimentFig}}
\end{figure}

Conjecturing the validity of  $\hat H_S$ with $\Omega_0^{eff}$ at larger $N$ as well,
we used   $\hat H_S$ to calculate the CFS for the  parameters of   Ref.~\cite{Campbell2009}. The results are shown in Fig.~\ref{ExperimentFig}.  Our model gives reasonable agreement within the experimental data  error bars  for
a range of scattering lengths
$a_{eg}^-=(100-500)a_0$ 
with $\Omega_0^{eff}= 4 \Omega_0$. 
The net effect of $\Omega_0^{eff}$ is to rescale the shift by a factor of four \footnote{For fixed pulse areas, an effective  bare Rabi frequency  renormalizes the CFS according to the following scaling law  $\delta \omega_{eg}(\Omega_0,\bar{U}, \delta )=\frac{\Omega_0}{\Omega_0^{eff}}\delta \omega_{eg}(\Omega_0^{eff},\bar{U} \frac{\Omega_0^{eff}}{\Omega_0},\delta \frac{\Omega_0^{eff}}{\Omega_0})$.}, which is justified by  the exchanging-modes  corrections discussed above and by the large (up to a factor of 5 
\cite{sebastian}) uncertainty in the experimental determination of the density. 
We also note that in Ref.~\cite{Campbell2009}  $\langle \Delta \Omega \rangle_T$  was inferred by fitting Rabi oscillations with a non-interacting model. However, the inset of  Fig.~\ref{ExperimentFig} shows that interactions modify Rabi oscillations and that the non-interacting model can underestimate $\langle \Delta \Omega \rangle_T$  by  a factor as large as three.
All these issues combined with the experimental uncertainty in pulse area complicate the 
determination of the magnitude of $a_{eg}^-$ from the current data. Nevertheless, our model  at least suggests  it to be positive,  $a_{eg}^->0$.

In addition to $a_{eg}^-$, there is another scattering length $a_{eg}^+$, which characterizes collisions between a $g$ and an $e$ atom. $a_{eg}^+$  collisions require a symmetric electronic state 
and consequently an anti-symmetric nuclear spin configuration. Thus the 
interaction Hamiltonian describing $g$-$e$ collisions beyond the polarized regime is  \cite{Gorshkov2009,Gorshkov20092}
\begin{equation}
\hat H_{int} = u_{eg}^s
\int d^3 \mathbf{r}\hat  \rho_e \hat \rho_g  + u_{eg}^a
\sum_{m m'} \int d^3 \mathbf{r} \hat \Psi^\dagger_{g m} \hat \Psi^\dagger_{e m'} \hat \Psi_{g m'} \hat \Psi_{e m}. 
\end{equation}
Here $u_{eg}^{s,a}= (u_{eg}^+ \pm u_{eg}^-)/2$, $s$ ($a$) stands for  symmetric (antisymmetric),  $m, m' \!=\!-I,\dots, I$ label the 
nuclear Zeeman levels, $u_{eg}^+= 4 \pi \hbar^2 a_{eg}^+/M$, and  $\hat \rho_{\alpha}(\mathbf{r}) = \sum_{m} \hat \Psi^\dagger_{\alpha {m}}(\mathbf{r}) \hat \Psi_{\alpha {m}}(\mathbf{r})$. 

We now propose how 
to estimate $a_{eg}^+$ experimentally. The method   is  identical to that of  Ref.~\cite{Campbell2009} except  that probe light polarization should be circular instead of linear. 
The idea is to Rabi interrogate 
an ensemble of 
$|e,m_0\rangle$ atoms with a circularly polarized probe 
driving  the   $|e,m_0\rangle-|g,m_0+1\rangle$ transition. If $a_{eg}^+=a_{eg}^-$, then $u_{eg}^a = 0$ 
and the dynamics of the system 
are identical  to the ones described above, except that one needs to identify $\{|e,m_0\rangle,|g,m_0+1\rangle\}$ as the 
spin basis. 
However, if $a_{eg}^+\neq a_{eg}^-$, 
the term proportional to $u_{eg}^a$  will populate $|e,m_0+1\rangle$ and $|g,m_0\rangle$. 
This issue  
can be overcome by applying an external magnetic field. 
If the applied magnetic field satisfies 
$B \mu_N \Delta g \gg   \langle \bar{U}^a \rangle_T$, 
with $\mu_N$ the nuclear magneton, $\Delta g$ the differential 
$g$-factor between $e$ and $g$ \cite{Boyd2006, Boyd2007}, and $ \langle \bar{U}^a \rangle_T$ the thermally averaged antisymmetric interaction, then the 
processes populating $|e,m_0+1\rangle$ and $|g,m_0\rangle$ will be energetically suppressed, and the dynamics will be identical to the ones in Ref.~\cite{Campbell2009} with CFS  
proportional to $u_{eg}^s$. By comparing  the CFS between the linearly and circularly polarized cases, one can in principle infer  
$a_{eg}^+$. 

In summary, fermionic clocks are sensitive to the CFS induced by
excitation inhomogeneities. The CFS  is sensitive to  pulse area, detuning, and interaction strengths, but if measured close to resonant transfer, it is in qualitative agreement with the  expected mean field expression. To  improve the experimental resolution of the CFS, better control over pulse area variations is required.  Rabi interrogation schemes can also be used 
to estimate  $a_{eg}^+$.

{\it Note added in proof:} While writing  this paper, we learned about a related  study of  interaction frequency shifts  in clock experiments \cite{Gibble2009}.

We gratefully acknowledge conversations with and feedback from J.~Ye, S.~Blatt, G.~Campbell, A.~Ludlow, P.~Julienne, and M.~Lukin. This work was supported by NSF.

\bibliography{./ref}

\end{document}